\documentclass[aps,pre,twocolumn, titlepage]{revtex4}

\usepackage{graphicx}
\usepackage{color}
\usepackage{dcolumn}
\usepackage{amsfonts}
\usepackage{bm}

\usepackage{epstopdf}

\begin{document}

\title{Depinning dynamics of two-dimensional dusty plasmas on a one-dimensional periodic substrate}

\author{W. Li$^1$, K. Wang$^1$, C. Reichhardt$^2$, C. J. O. Reichhardt$^2$, M. S. Murillo$^3$, and Yan Feng$^1 \ast$}
\affiliation{
$^1$ Center for Soft Condensed Matter Physics and Interdisciplinary Research, School of Physical Science and Technology, Soochow University, Suzhou 215006, China\\
$^2$ Theoretical Division, Los Alamos National Laboratory, Los Alamos, New Mexico 87545, USA\\
$^3$ Department of Computational Mathematics, Science and Engineering, Michigan State University, East Lansing, Michigan 48824, USA\\
$\ast$ E-mail: fengyan@suda.edu.cn}

\date{\today}

\begin{abstract}

We investigate the depinning dynamics of two-dimensional dusty plasmas (2DDP) driven over one-dimensional periodic substrates (1DPS) using Langevin dynamical simulations. We find that, for a specific range of substrate strengths, as the external driving force increases from zero, there are three different states, which are the pinned, the disordered plastic flow, and the moving ordered states, respectively. These three states are clearly observed using different diagnostics, including the collective drift velocity, static structural measures, the particle trajectories, the mean-squared displacements, and the kinetic temperature. We compare the observed depinning dynamics here with the depinning dynamics in other systems.

\end{abstract}

\maketitle

\section{Introduction}
There are a wide variety of particle-like systems
that, when coupled to some form
of underlying substrates, exhibit depinning under an applied drive~\cite{Reichhardt:2016}.
These include
vortices in type-II superconductors~\cite{Bhattacharya:1993,Koshelev:1994,Pardo:1998,Olson:1998},
colloids~\cite{Pertsinidis:2008,Bohlein:2012,Tierno:2012}, Wigner crystals~\cite{Williams:1991,Cha:1998},
and pattern forming systems \cite{Reichhardt:2003,Sengupta:2010}.
For collections of interacting
particles, the depinning can be either elastic~\cite{Reichhardt:2016,Scala:2012},
in which all the particles maintain the same neighbors
and retain topological order during depinning,
or plastic, where neighbors
exchange with each other
and topological disorder can be generated \cite{Reichhardt:2016,Fily:2010}.
For systems that exhibit plastic depinning,
at higher drives beyond depinning,
further transitions can occur from a plastically moving state
into a moving crystal \cite{Koshelev:1994,Pardo:1998,Olson:1998,Doussal:1998}
or moving smectic state \cite{Pardo:1998,Olson:1998,Balents:1998,Reichhardt:2001}
which are associated with the
emergence of hexatic ordering.

Another class of system that
can be modeled as an assembly of
interacting particles is complex or dusty plasmas \cite{Chu:1994,Thomas:1994,Morfill:2009,Fortov:2005,Piel:2010,Bonitz:2010,Merlino:2004,Feng:2008,Thomas:2004}, which refer to a partially ionized gas containing highly charged micron-sized dust particles. Under typical laboratory conditions, these dust particles are negatively charged to $\approx - 10^4 e$, so that they are strongly coupled, exhibiting collective solidlike~\cite{Feng:2008,Hartmann:2014} and liquidlike~\cite{Chan:2007,Feng:2010} behaviors. These dust particles can be levitated and confined in the plasma sheath by an electric field, so that they can self-organize into a single-layer suspension with the negligible out-of-plane motion, called a two-dimensional dusty plasma (2DDP)~\cite{Feng:2011}.
Although dusty plasmas have been studied for about two decades,
it was only recently
proposed to couple a dusty plasma to
a one-dimensional periodic substrate (1DPS)~\cite{Li:2018,Wang:2018}.
Distinct features in the phonon spectra~\cite{Li:2018} and self-diffusion~\cite{Wang:2018}, as well as the melting transitions~\cite{Wang:2018}, appear in the presence of the substrate
which are absent in a substrate-free 2DDP system.

A natural next question to address is whether
dusty plasmas also exhibit a depinning
transition under an applied drive.
Building upon our previous work on dusty plasmas coupled to
1DPS~\cite{Li:2018,Wang:2018},
here we study the driven dynamics of dusty plasmas coupled to 1DPS for
varied substrate coupling and drives.
We find that this system exhibits plastic depinning,
with a disordered mixing of the particles at the
depinning threshold correlated with the proliferation of
non-sixfold coordinated particles.
At the highest drives,
the system dynamically reorders into a moving triangular crystal with the hexagonal symmetry,
and interestingly the
moving structure is more ordered than the system with pinning at zero drive.
For weaker substrate strengths, the
depinning is continuous,
but for stronger substrates the depinning is a discontinuous
transition.  This is in
contrast to the depinning of overdamped particles
on a 1D substrate where the depinning threshold
is always continuous.
The crossover to discontinuous depinning is consistent with predictions for
the depinning of systems in which inertial or overshoot
effects are important~\cite{Schwarz:2001}. Our results indicate that
dusty plasmas represent
another class of system that
can exhibit depinning and nonequilibrium driven sliding
phases when coupled to a periodic substrate.

This paper is organized as follows. In Sec.~II, we briefly introduce our Langevin dynamical simulation method to mimic 2DDP interacting with 1DPS and a driving force. In Sec.~III, we present our results on the structural and dynamical properties, as functions of the external driving force. Finally, we provide a brief summary.

\section{Simulation method}

Two-dimensional dusty plasmas systems are usually characterized using the coupling parameter $\Gamma$ and the screening parameter $\kappa$~\cite{Ohta:2000,Sanbonmatsu:2001}. Here,
\begin{equation}\label{Parameter}
{ \Gamma = Q^2/(4 \pi \epsilon_0 a k_B T), }
\end{equation}
where $Q$ is the particle charge, $T$ is the particle kinetic temperature, $a = (n\pi)^{-1}$ is the Wigner-Seitz radius~\cite{Kalman:2004}, and $n$ is the areal number density. The screening parameter $\kappa = a/\lambda_{D}$ indicates the length scale of the space occupied by one dust particle over the Debye screening length $\lambda_D$. In addition to the value of $a$, we use the lattice constant $b$ to normalize the length. For a 2D defect-free triangular lattice, $b = 1.9046a$.

We use Langevin dynamical simulations to study the depinning dynamics of 2DDP on a 1DPS with external driving force. The equation of motion for dust particle $i$ is
\begin{equation}\label{LDE}
{	m \ddot{\bf r}_i = -\nabla \Sigma \phi_{ij} - \nu m\dot{\bf r}_i + \xi_i(t)+{\bf F}_s + {\bf F}_d. }
\end{equation}
Here, the particle-particle interaction force, given by a Yukawa or screened Coulomb potential, is  $-\nabla \Sigma \phi_{ij}$, with $\phi_{ij} = Q^2 {\rm exp}(-r_{ij} / \lambda_D) / 4 \pi \epsilon_0 r_{ij}$, where $r_{ij}$ is the distance between particles $i$ and $j$, and $-\nu m\dot{\bf r}_i$ is the frictional gas drag~\cite{Liu:2003}. The Langevin random kicks term $\xi_i(t)$ is assumed to have a Gaussian distribution with a width that is related to the desired target temperature, according to the fluctuation-dissipation theorem~\cite{Feng:2008,Feng2:2008,Kalman2:2004}. The force ${\bf F}_s$ comes from the 1DPS, as shown in Fig.~\ref{fig:Distribution}, which has the sinusoidal form
\begin{equation}\label{1DPS}
{	U(x) = U_0 \cos(2 \pi x/w).}
\end{equation}
The resulting force is ${\bf F}_s = - \frac {\partial U(x)}{\partial x} \hat{\bf x} = (2\pi U_0/w)\sin(2\pi x/w) \hat{\bf x}$. Here, $U_0$ is the substrate strength in units of $E_0 = Q^2/4\pi\epsilon_0 a $ and $w$ is the width of the substrate in units of $b$, respectively. The last term on the right-hand side of Eq.~(\ref{LDE}) ${\bf F}_d = F_d \hat{\bf x}$ is the applied external driving force, in units of $Q^2/4\pi\epsilon_0 a^2 $.

Our simulation includes 1024 particles in a rectangular box of dimensions $61.1a \times 52.9a$ with the periodic boundary conditions. We fix $\Gamma = 1000$ and $\kappa = 2$ as the conditions of 2DDP to reduce the effect of temperature on the depinning behavior, since the melting point of 2DDP is $\Gamma = 396$ when $\kappa = 2$~\cite{Hartmann:2005}. Since the simulated size is $61.1a \approx 32.07b$ in the $x$ direction, we choose $w = 1.002b$ corresponding to 32 full potential wells to satisfy the periodic boundary conditions. For the substrate strength, we specify three values of $U_0 = 0.0025E_0$, $0.005E_0$, and $0.01E_0$, where the unit is $E_0 = Q^2/4\pi \epsilon_0 a$. We increase the external driving force $F_d$ along the horizontal direction $x$ and measure the drift velocity $V_x = N^{-1} <\sum_{i=1}^{N}{\bf v}_i \cdot \hat {\bf x}> $, where $N$ is the number of particles, at each driving force increment.

For each simulation run, we integrate Eq.~(\ref{LDE}) to record the particle positions and velocities in $10^7$ simulation steps using a time step of $0.003\omega^{-1}_{pd}$ for the latter data analysis. Here, ${\omega}_{pd} = (Q^2/2\pi\epsilon_0 m a^3)^{1/2}$ is the nominal dusty plasma frequency. We specify the frictional gas damping as $ \nu / {\omega}_{pd} = 0.027$, comparable to typical dusty plasma experiments~\cite{Feng:2011}. Other simulation details are the same as those in~\cite{Wang:2018,Li:2018}. In addition, we also perform a few test runs for a larger system size of  $N = 4096$ dust particles, and we do not find any substantial differences in the results reported here.

\section{Results and discussions}

\subsection{Pinning and depinning}

\begin{figure*}[htb]
	\centering
	\includegraphics{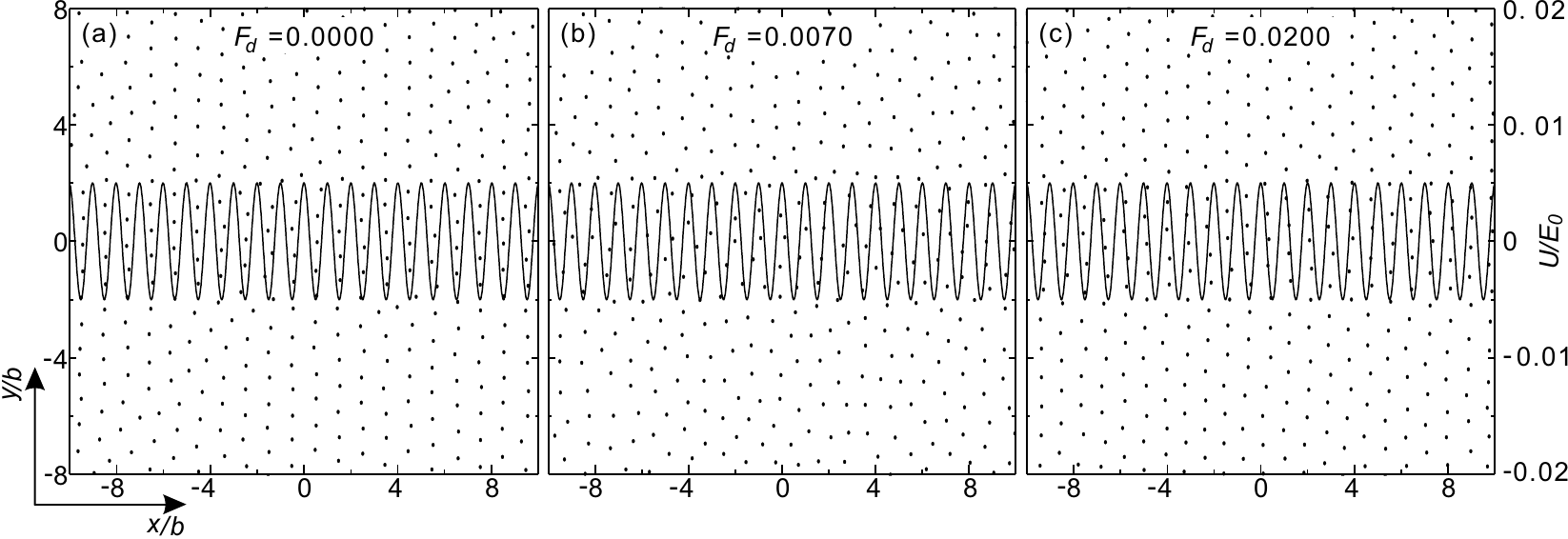}
	\caption{\label{fig:Distribution}
          Snapshots of particle positions (dots) and the locations of the 1DPS (curve) in a 2D Yukawa crystal with $\Gamma=1000$ and $\kappa=2$ under different external driving forces for systems with the 1DPS $U(x) = U_0 \cos(2\pi x/w)$ of
strength $U_0 = 0.005 E_0$ and period $w = 1.002 b$.
(a) At $F_d=0$, all particles are pinned at the bottom of each substrate potential well, forming a pinned state composed of 1D or quasi-1D chains.
(b) At $F_d=0.007$, the particle distribution is disordered, which is independent of the spatial distribution of the potential wells.
(c) At $F_d=0.02$, the particles are distributed nearly in a triangular lattice, which is also independent of the spatial distribution of the potential wells.
        }
\end{figure*}

Snapshots of particle positions from our simulations present the particle distribution of the 2DDP on the 1DPS for typical values of the external driving force $F_d$, as shown in Fig.~\ref{fig:Distribution}. Here, the conditions of the 2DDP and 1DPS are fixed, where $\Gamma = 1000$, $\kappa=2$, $U_0 = 0.005E_0$ and $w = 1.002b$. The only modified variable is the driving force, which is $F_d = 0$, $0.007$, and $0.02$ for Figs.~\ref{fig:Distribution}(a), (b), and (c), respectively.

When there is no external driving force, all particles are pinned at the bottom of each potential well of the substrate, forming several 1D or quasi-1D chains, as shown in Fig.~\ref{fig:Distribution}(a). Similar results have been reported in~\cite{Li:2018, Wang:2018}. Figure~\ref{fig:Distribution}(b) presents the particle distribution with the external driving force of $F_d = 0.007$. The particle distribution is strongly disordered, which is independent of the locations of the substrate potential wells. We attribute this result to the competition between the external driving force and the constraint imposed by the 1DPS, and probably only some particles can overcome the constraint from the 1DPS. Figure~\ref{fig:Distribution}(c) presents the particle distribution with the external driving force of $F_d = 0.02$. The particles are arranged in a nearly triangular lattice, which is probably due to the fact that the external driving force is large enough to totally move all particles across the constraint of 1DPS as a whole object. Thus, the particle arrangement is independent of the 1DPS when the external driving force is large enough.

\begin{figure}[htb]
    \centering
    \includegraphics{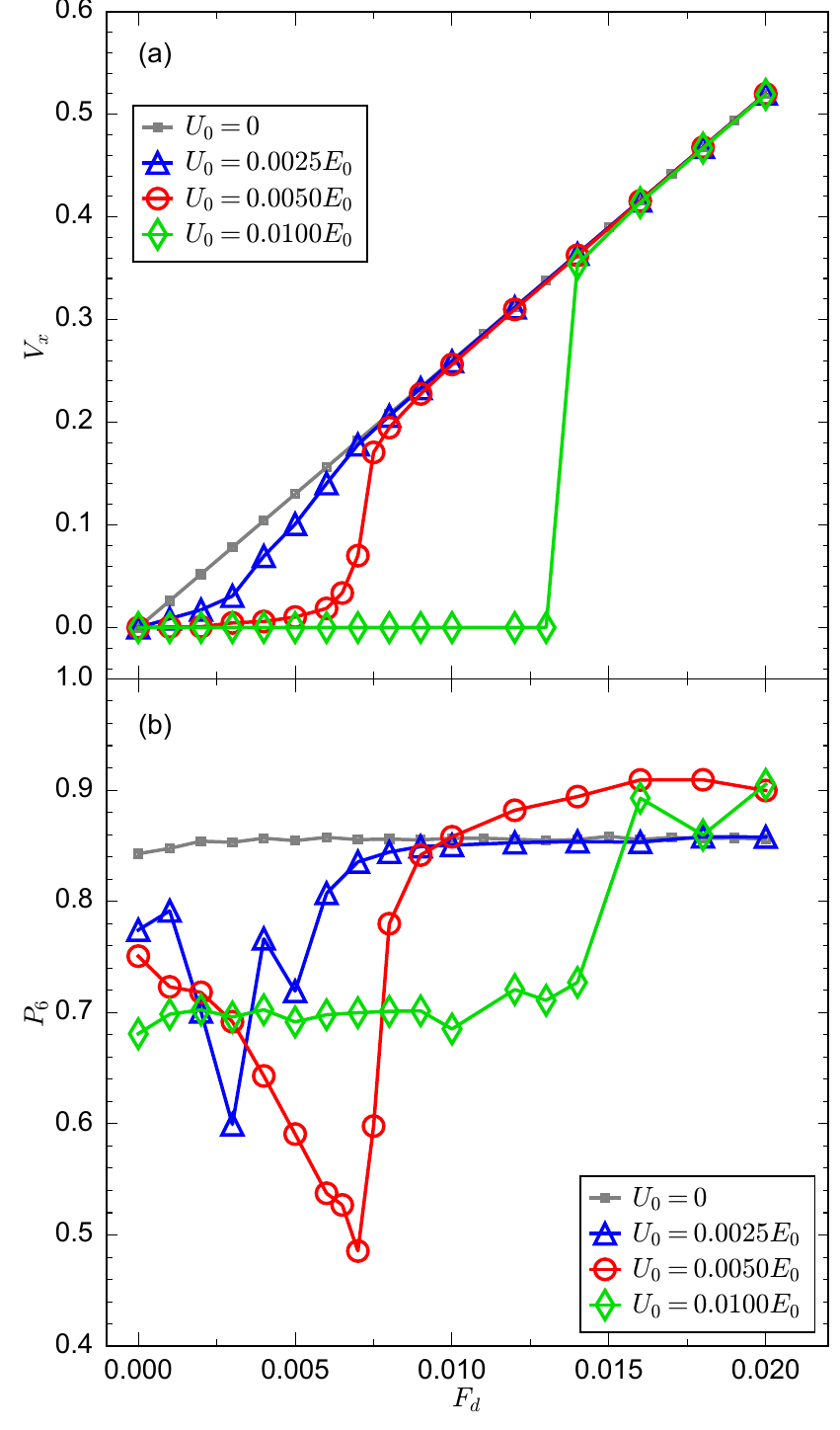}
    \caption{\label{fig:Dynamics}
(a) The collective drift velocity $V_x$ vs external driving force
$F_d$ for substrate strengths $U_0=0.0025E_0$,
0.005$E_0$, and $0.01E_0$, respectively.
(b) The corresponding fraction of sixfold coordinated
particles~\cite{Reichhardt:2005} $P_6$ vs $F_d$ for the same
substrate strengths.
As $F_d$ increases from zero, for the $U_0=0.0025E_0$ and $U_0=0.005E_0$
samples
we observe three states: pinned, disordered plastic flow,
and a moving ordered lattice.
For comparison, the lines with small squares
indicate the response
for particles sliding freely without a substrate, $U_0=0$.
    }
\end{figure}

In Fig.~\ref{fig:Dynamics}(a) we plot
the collective drift velocity $V_x$ versus the applied driving force $F_d$, while in
Fig.~\ref{fig:Dynamics}(b) we show the corresponding
fraction of sixfold coordinated particles $P_6$ versus $F_d$
for four different values of $U_0$.
Here, $V_x$ is the collective drift velocity along the direction of the driving force,
$V_x = N_d^{-1} \langle \sum_{i=1}^{N_d}{\bf v}_i \cdot \hat {\bf x}\rangle$, while the fraction of sixfold coordinated particles $P_6$ is
obtained \cite{Reichhardt:2005} from $P_6 = N_d^{-1}\langle \sum_{i=1}^{N_d}\delta(6-z_i)\rangle$,
where $z_i$ is the coordination number of particle $i​$
obtained from the Voronoi construction.
For a perfect triangular lattice, $P_6 = 1.0$, while the value of $P_6$ is
reduced in a more disordered system.

For the intermediate substrate strengths of
$U_0=0.0025E_0$ and 0.005$E_0$, we
find evidence for three dynamic phases, each of which
has characteristic signatures in both $V_x$ and $P_6$.
The first state appears for small drives, when the driving force is
small and the collective drift velocity is nearly zero.  In this
pinned state \cite{Reichhardt:2016}, the external driving force is not large
enough to overcome the attraction of the substrate, so all particles
remain nearly stationary around their original equilibrium locations and
the lattice structure is still highly ordered, as shown in Fig.~\ref{fig:Distribution}(a).
At very large drives, $F_d>0.01$, the collective drift velocity
increases linearly with $F_d$, as shown in Fig.~\ref{fig:Dynamics}(a),
and the structure is highly ordered, as indicated by the fact that $P_6>0.8$
in Fig.~\ref{fig:Dynamics}(b).  In this moving ordered state
\cite{Reichhardt:2016}, the external driving force is so large that all of the
particles move over the 1DPS as a stiff solid object, as illustrated
in Fig.~\ref{fig:Distribution}(c).
For intermediate drives of $0.004 < F_d < 0.007$ for $U_0=0.005E_0$ or
$0.002 < F_d < 0.006$ for $U_0=0.0025E_0$, the collective drift
velocity increases relatively steeply, as shown in
Fig.~\ref{fig:Dynamics}(a), and the structure is more disordered,
as indicated by the low value of $P_6$ in Fig.~\ref{fig:Dynamics}(b).
This is the typical disordered plastic flow state \cite{Reichhardt:2016}.
A jump in $V_x$ can arise
at the transitions
either because more particles begin moving suddenly, 
or because there is a change
from a disordered plastic flow state
or a pinned state to a moving ordered state.
These dynamic phases have not been observed previously in dusty plasmas,
although they are similar to
results reported for colloids
driven over a 1DPS
~\cite{Reichhardt:2005} and to phases that have been found in other
systems
~\cite{Koshelev:1994,Thorel:1973,Shi:1991,Reichhardt:2015}.

The existence of three dynamical states depends on the value of the
substrate strength. For extremely strong substrates, such as
$U_0=0.01E_0$, the disordered plastic flow state is destroyed and the
particles undergo an abrupt depinning transition directly into a moving
ordered state, as shown by the rapid increase in both $V_x$ and $P_6$
in Fig.~\ref{fig:Dynamics}. If the substrate is absent, as shown for
$U_0=0$ in Fig.~\ref{fig:Dynamics}, the drift velocity always increases
linearly with $F_d$ and the structure is always highly ordered, so that only
the moving ordered state exists.

From our results of Fig.~\ref{fig:Dynamics} above, it is clear that the
depinning is continuous for weaker substrate strengths. However, for stronger substrates, the depinning is a discontinuous transition.  This feature is completely different from the depinning of overdamped colloidal particles
on a 1D substrate, where the depinning threshold
is always continuous.
In~\cite{Schwarz:2001}, it is predicted that the inertial or overshoot
effects are important in the depinning dynamics. 
Our observation of a discontinuous depinning here is consistent with the prediction in~\cite{Schwarz:2001}. 
For our simulated 2DDPs, the existence of the inertial term, as well as the underdamped motion of particles, modifies the  depinning behavior and produces the discontinuity. 

Regardless of the value of the substrate strength, for sufficiently large
driving forces, the collective drift velocity increases linearly with the driving force 
at a fixed slope, as shown in Fig.~\ref{fig:Dynamics}(a).
In our simulated 2DDP, the gas damping is specified as
$\nu=0.027 \omega_{pd}$, in agreement with typical experimental values
\cite{Feng2:2008}.  In the moving ordered state, the particles reach a
steady state velocity, in which the driving force is balanced by the frictional
gas drag, ${\bf F}_d = -\nu m\dot{\bf r}_i$. The relative motion of
particles with respect to each other is negligible compared to the collective drift
motion, so the speed of each individual particle is nearly identical to
the collective drift velocity, $\dot{\bf r}_i \approx V_x \hat {\bf x}$.
Thus, in the moving ordered state, the slope of the drift velocity
$V_x$ as a function of driving force $F_d$ is given by
$F_d/V_x = -\nu m\dot{\bf r}_i/V_x \cdot \hat{\bf x} = -m\nu$,
meaning that it is purely determined by the gas damping value in our simulations.

\subsection{Three phases from various diagnostics}

\begin{figure}[htb]
	\centering
	\includegraphics{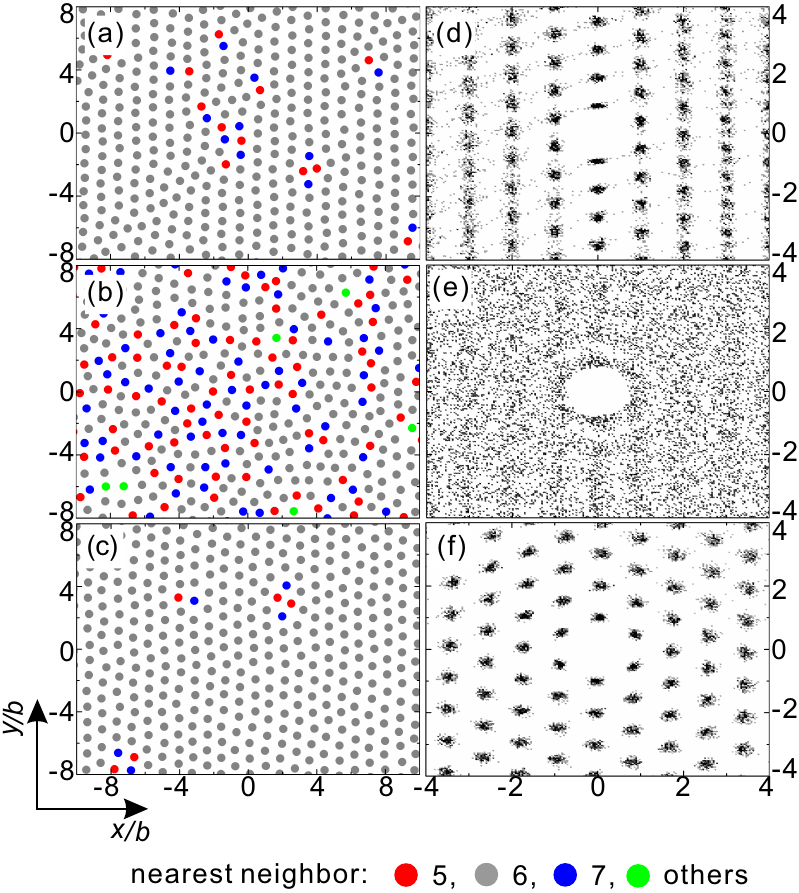}
	\caption{\label{fig:Phase}
Particle configurations (left column) and 2D distribution functions
$G_{xy}$ (right column) for the 2DDP system with $\Gamma=1000$ and $\kappa=2$
on a 1DPS with $U_0=0.005E_0$ at drives of (a, d) $F_d=0$ in the pinned state,
(b, e) $F_d=0.007$ in the disordered plastic flow state,
and (c, f) $F_d=0.02$
in the moving ordered state.
In the left column, particles are colored according
to their coordination number of neighbors as marked in the legend at the bottom.
          }
\end{figure}

To further verify the structural measurement results of $P_6$ in
Fig.~\ref{fig:Dynamics}(b), we image the locations of the
particles and calculate the 2D distribution function
$G_{xy}$, as shown in Fig.~\ref{fig:Phase} for a system with
$U_0=0.005E_0$.
In Fig.~\ref{fig:Phase}(a, d) at $F_d=0$, most of the particles have
six nearest neighbors, and $G_{xy}$ shows strong ordering, consistent
with the high value of $P_6$ found in Fig.~\ref{fig:Dynamics}(b)
for the pinned state.
At $F_d=0.007$ in Fig.~\ref{fig:Phase}(b, e), both the particle
configuration and $G_{xy}$ indicate strong disordering of the
particle positions. There are also a large number of particles that
do not have six neighbors, as shown in Fig.~\ref{fig:Phase}(b).
This agrees well with the low value of $P_6$ in Fig.~\ref{fig:Dynamics}(b)
for the disordered plastic flow state.
When $F_d=0.02$, as in Fig.~\ref{fig:Phase}(c, f), the particle configuration
is ordered again, most particles have six nearest neighbors, and
$G_{xy}$ clearly shows strong ordering, which is again consistent with the
high value of $P_6$ in Fig.~\ref{fig:Dynamics}(b) in the moving
ordered state. The structure of the
moving ordered state is more ordered than that of the pinned state at zero driving.

\begin{figure*}[htb]
	\centering
	\includegraphics{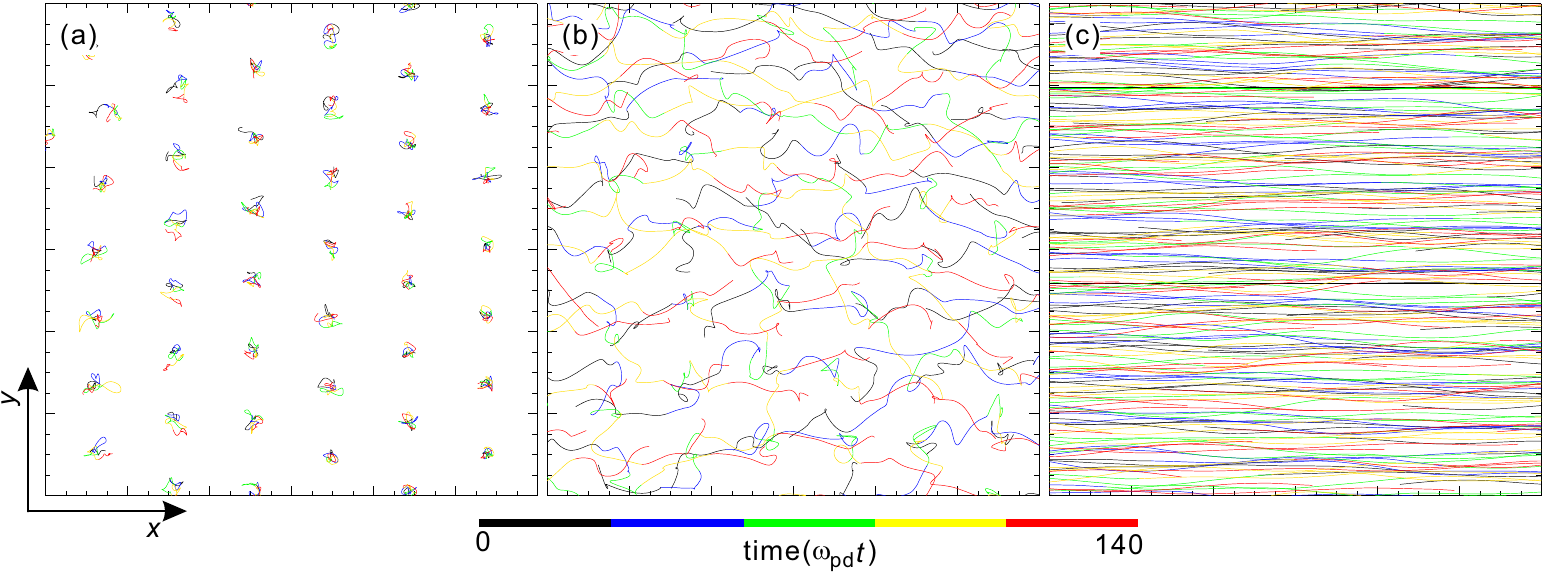}
	\caption{\label{fig:Trajectory}
Typical particle trajectories 
from the simulated 2DDP
for the same conditions as in Fig.~\ref{fig:Phase}
under different constant external driving forces.
(a) At $F_d= 0$ in the pinned state, all of the particles are pinned around
their equilibrium locations and undergo only the caged motion.
(b) At $F_d = 0.007$ in the disordered plastic flow state, some of the
particles can escape from the cages and move out of the potential minima.
These particles immediately become trapped in new cages or potential wells,
from which they can once again escape, resulting in trajectories that are
disordered along both the $x$ and $y$ directions.
(c) At $F_d = 0.02$ in the moving ordered state, the driving force is so
large that the particles can easily cross over the potential barriers.
As a result, the particles move predominantly parallel to the driving
direction (the $x$ direction), with negligible motion in the $y$
direction. These trajectories are obtained from $\approx 4\%$ of 
the simulation box over a time period representing $\approx 0.5\%$ of the full duration of the simulation.
}
\end{figure*}

We next consider the particle trajectories in each of the three
dynamical phases.
In Fig.~\ref{fig:Trajectory}
we illustrate typical trajectories of the simulated 2DDP (with
$\Gamma = 1000$ and $\kappa=2$)
on a 1DPS with $U_0 = 0.005E_0$
at
$F_d = 0$ in the pinned state, $F_d=0.007$ in the
disordered plastic flow state, and $F_d=0.02$ in the moving
ordered state, respectively.
Color represents time, and only $\approx 4\%$ of the simulated region and $\approx 0.5\%$ of the simulation duration are shown here.
In the pinned state of Fig.~\ref{fig:Trajectory}(a),
all of the particles are trapped at their equilibrium locations
and can only undergo the caged motion,
with no particles able to escape from its cage or potential well.
In the disordered plastic flow state of Fig.~\ref{fig:Trajectory}(b),
the particle trajectories
are very disordered.
At one instant in time, some particles are moving rapidly while other
particles remain stationary, but at a later time, the particles switch roles,
so that some of the moving particles are now stationary while some of the
trapped particles are now moving.
The particles do not move strictly along the driving or $x$ direction, but
also undergo considerable motion in the perpendicular or $y$ direction.
This is a consequence of the importance of the particle-particle interactions.
In order for one particle to escape from a substrate potential well,
its neighboring particle inside the same potential well must shift out of the
way by moving along the $y$ direction, parallel to the orientation of the
substrate minimum, as shown in Fig.~\ref{fig:Trajectory}(b).
In the moving ordered state of Fig.~\ref{fig:Trajectory}(c), the external driving force is large enough that
all of the particles can easily cross the potential barriers simultaneously,
allowing
each particle to maintain
the same neighbors as it moves~\cite{Moon:1996}.
The trajectories clearly
indicate that all of the particles move almost completely parallel to the
driving force direction, with negligible motion in the perpendicular
or $y$ direction.

As described above, in our simulations of 2DDP over 1DPS, when the driving force increases, we observe three phases: the pinned state, the disordered plastic flow state, and the moving ordered state. Similar phases were also observed in other physical systems or theoretical models~\cite{Koshelev:1994, Reichhardt:2002, Granato:2011}. Based on 2D simulations of vortex lattices~\cite{Koshelev:1994}, Koshelev and Vinokur mapped out the dynamical phase diagram of the vortex system, which contains a pinning phase, a plastic flow phase, and a moving crystal. Using 2D simulations of colloids~\cite{Reichhardt:2002}, Reichhardt and Olson found three dynamical phases consisting of pinned, plastic flow, and elastic flow. From 2D simulations of a phase-field-crystal model~\cite{Granato:2011}, Granato et al. obtained a dynamical phase diagram containing a pinned amorphous glass, plastic flow, and a moving smectic glass.

\subsection{Dynamical properties of different directions}

\begin{figure}[htb]
    \centering
    \includegraphics{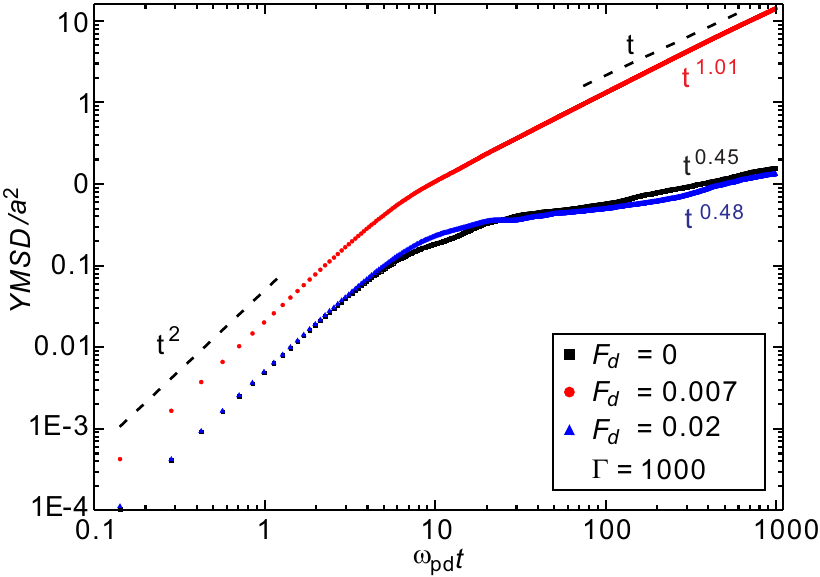}
    \caption{\label{fig:MSDY}
Mean-squared displacement (MSD) calculated from the motion in the $y$ direction, marked as YMSD here, for
the simulated 2DDP
under the same conditions as in Fig.~\ref{fig:Phase}
at different constant external driving forces of $F_d = 0.0$ (squares), $0.007$ (circles),
and $0.02$ (triangles).
For both the initial ballistic motion at early times and the later
diffusive motion at longer times, we find that,
as the driving force increases from zero, the
YMSD always increases first and then decreases again.
Furthermore, for the later diffusive motion at $F_d=0.007$ in the
disordered plastic flow state,
we find $YMSD \propto t^{1.01}$.
This behavior is distinct from that found in the
pinned state,
${\rm YMSD} \propto t^{0.45}$,
and in the moving ordered state, ${\rm YMSD} \propto t^{0.48}$.
The simulation conditions of our 2D Yukawa crystal are $\Gamma = 1000$ and $\kappa =2$.
      }
\end{figure}

To study the dynamical properties of 2DDP on 1DPS perpendicular to the external driving force, we calculated the mean-squared displacement due to motion in the $y$ direction only (YMSD), as described in Ref.~\cite{Wang:2018}.
Figure~\ref{fig:MSDY} shows
the YMSD
for the same sample from Fig.~\ref{fig:Phase}
at $F_d = 0$, $0.007$, and $0.02$, respectively.
For the longer time diffusive motion,
the MSD can be fit to the expression ${\rm MSD} = D t^\alpha$,
where the exponent $\alpha$ reflects the diffusion properties.
A value $ \alpha = 1 $ indicates that normal diffusion
is occurring, while $ \alpha > 1 $ and $\alpha < 1$
correspond to super- and subdiffusion, respectively.
We find that
the pinned and moving ordered states have nearly the same YMSD behavior,
while the YMSD for the disordered plastic flow state is completely different.
As the external driving force increases from zero,
the YMSD first increases and then decreases,
both for the initial ballistic motion~\cite{Liu:2008} and
for the later diffusive motion.
We fit the YMSD
for the later diffusive motion
to
${\rm YMSD} = D t^\alpha$
for the time interval $100< \omega_{pd}t <990$,
and obtain exponents of $\alpha=0.45$ in the pinned state,
$\alpha=1.01$ in the disordered plastic flow state,
and $\alpha=0.48$ in the moving ordered state, as shown in Fig.~\ref{fig:MSDY}.
Clearly, in the pinned and moving ordered states,
the particles undergo the subdiffusive motion in the $y$ direction,
while in the disordered plastic flow state,
the particle motion along $y$ is
nearly the normal diffusion.
For both the pinned and moving ordered states,
the $y$-direction motion of the particles
is constrained by the neighboring particles in the lattice,
while in the plastic flow state, the lattice is disordered and this
constraint is lost, permitting the particles to move more freely
along the $y$ direction.

\begin{figure}[htb]
    \centering
    \includegraphics{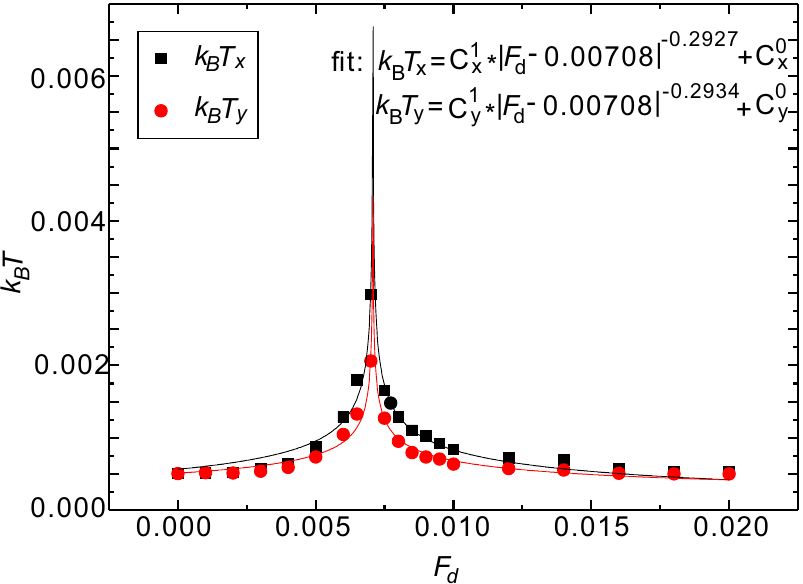}
    \caption{\label{fig:KinTem}
The kinetic temperature $k_BT_x$ (black squares) and $k_BT_y$ (red circles),
calculated from the motion in the $x$ and $y$ directions,
versus $F_d$ for the simulated 2DDP
at the same conditions as in Fig.~\ref{fig:Phase}.
The center-of-mass motion has been removed in calculating the kinetic temperature, so that only the fluctuations of the particle velocities are considered. 
As $F_d$
increases from zero,
both $k_BT_x$ and $k_BT_y$ first simultaneously increase and then decrease.
The initial increase is correlated with
the transition
from the pinned state to the disordered plastic flow state,
while the
decrease at higher drives is correlated with
the transition
from the disordered plastic flow state to the moving ordered state.
In addition, in the disordered plastic flow state,
we find that the kinetic temperatures are
anisotropic, with $k_BT_x \textgreater k_BT_y$,
due to the wider range of the $x$ direction
velocity distribution in this phase,
while in the pinned and moving ordered state, the anisotropy is
lost and
$k_BT_x \approx k_BT_y$.
To quantify the variation
in the kinetic temperature,
we fit it to the expression
$k_BT = C_1*|F_d-C_2|^\beta + C_0 $.
Curves with the fitting exponents of
$\beta = -0.2927$ and $-0.2934$ for $k_BT_x$ and $k_BT_y$, respectively,
are presented.
      }
\end{figure}

We plot
the kinetic temperatures $k_BT_x$ and $k_BT_y$,
calculated from the motion in the $x$ and $y$ directions
for the same system as in
Fig.~\ref{fig:Phase}, as functions of $F_d$
in Fig.~\ref{fig:KinTem}.
To obtain these kinetic temperatures, we use the equation
$k_BT = m\langle \sum^N_{i=1} ({\bf v_i} - \overline {\bf v})^2\rangle/2$,
where
the collective drift velocity $\overline {\bf v}$ is removed. As the external force increases from zero,
both $k_BT_x$ and $k_BT_y$
increase
substantially, and then decrease back to low values.
The initial increase occurs as the system transits
from the pinned state to the disordered plastic flow state,
while the decrease is associated with the transition
from the disordered plastic flow state
to the moving ordered state.
In the disordered plastic flow state we find an anisotropic
kinetic temperature
with $k_BT_x > k_BT_y$, as shown in Fig.~\ref{fig:KinTem}.
This is likely due to the fact that the $x$ direction velocity distribution
is wider in the disordered plastic flow state.

We find that the kinetic temperatures $k_BT_x$ and $k_BT_y$ can be
fit well
to the expression
$k_BT = C_1*|F_d-C_2|^\beta+C_0 $, and the resulting fitting curves are shown in
Fig.~\ref{fig:KinTem}.
The fitting parameter of $C_2$ corresponds to
the peak of the kinetic temperature, i.e., the most disordered plastic flow state, which is $C_2 = 0.00708$ from our fitting.
Fits of the exponent $\beta$ give $\beta = -0.2927$ and $-0.2934$ for $k_BT_x$ and $k_BT_y$, respectively. The significance of these exponent values is unclear, but the fluctuations in the velocity are largest as the driving force passes through the most disordered plastic flow state near $F_d = 0.00708$.

\section{Summary}

In summary, we study the depinning dynamics of a 2DDP on a 1DPS under a driving force using Langevin dynamical simulations. For a range of substrate strengths, we find three different dynamical phases as the external driving force increases from zero, which are a pinned state, a disordered plastic flow state, and a moving ordered state. Using different structural and dynamical diagnostics, such as the collective drift velocity, static structural measures, particle trajectories, mean-squared displacements, and the kinetic temperature, we quantify the differences between these three states. Such dynamical phases have not previously been observed in dusty plasmas. We compare our observations of the depinning dynamics of dusty plasmas with depinning in other previously studied systems.

\acknowledgements
Work in China was supported by the National Natural Science Foundation of China under Grant No. 11875199, 11505124, the 1000 Youth Talents Plan, startup funds from Soochow University, and the Priority Academic Program Development (PAPD) of Jiangsu Higher Education Institutions. Work at LANL was carried out under the auspices of the NNSA of the U.S. DOE under Contract No. DE-AC52-06NA25396.

\end{document}